\documentclass{elsart}    

\usepackage{graphicx}          

\usepackage{subfigure}
\usepackage{amsmath}
\usepackage{balance}

\begin{document}

\begin{frontmatter}

\title{Minimum-Time Cavity Optomechanical Cooling\thanksref{footnoteinfo}} 

\thanks[footnoteinfo]{The material of this paper has not been presented at any conference.}

\author[1]{Dionisis Stefanatos}
\ead{dionisis@post.harvard.edu}
\address{Hellenic Army Academy, Vari, Athens 16673, Greece}
\thanks[1]{Tel.: +30-697-4364682; Fax: +30-26740-23266.}


\begin{abstract}                          

Optomechanical cooling is a prerequisite for many exotic applications promised by modern quantum technology and it is crucial to achieve it in short times, in order to minimize the undesirable effects of the environment. We formulate cavity optomechanical cooling as a minimum-time optimal control problem on anti-de Sitter space of appropriate dimension and use the Legendre pseudospectral optimization method to find the minimum time and the corresponding optimal control, for various values of the maximum coupling rate between the cavity field and the mechanical resonator. The present framework can also be applied to create optomechanical entanglement in minimum time and to improve the efficiency of an optomechanical quantum heat engine.


\end{abstract}

\end{frontmatter}

\section{Introduction}

Quantum optomechanics is a rapidly growing field which investigates the quantum aspects of the interaction between light and mechanical motion \cite{Meystre13}. The mechanical effects of light originate from the fact that light carries momentum which gives rise to radiation pressure forces, we mention for example the experimental spacecraft IKAROS, whose flight is mainly powered by solar radiation pressure acting on a square solar sail with 20 $m$ diameter \cite{Mori09}. The large size of the solar sail indicates a drawback of optical forces exerted on macroscopic objects: they are very weak. In order to bypass this problem, optical cavities, formed by one fixed and one movable mirror attached to a spring, are used in the laboratory. The light trapped inside the cavity is resonantly amplified, enhancing thus the optical force acting on the movable mirror.

In the recent years, the field of cavity optomechanics has witnessed an unprecedented development with a wide range of applications \cite{Aspelmeyer14}. In the area of high-precision measurements, cavity optomechanics promises the detection of tiny forces, masses, displacements and accelerations, close to the fundamental limits imposed by quantum mechanics. In the field of quantum information processing, optomechanical devices can be used to convert solid-state ``storage" qubits into photonic ``communication" qubits, and also to create entanglement between macroscopic objects (for example the movable mirror) and the electromagnetic field in the cavity. From the last application it is also obvious that cavity optimechanics provides the ideal platform for tests of fundamental quantum physics in a macroscopic range of parameters like size and mass.

At the heart of these exotic applications lies the problem of cooling the mechanical resonator (movable mirror) to its ground state, so the full advantage of the quantum behavior is exploited \cite{Liu13}. Various techniques from quantum control \cite{Dong10} have been suggested to overcome this challenge. Measurement-based feedback was initially proposed \cite{Mancini98} and implemented experimentally \cite{Cohadon99}, where the mirror displacement is continuously measured, using the phase shift that induces on the output radiation field, and the produced signal is fed back into the cavity to generate a damping force opposing mirror motion. Around the same time, the use of Kalman filter was introduced \cite{Doherty99}, to obtain a better estimate of the mirror position; it was recently demonstrated in the laboratory \cite{Wieczorek15}. Later, using the theoretical tools developed in \cite{James08,Gough09,Nurdin09}, coherent feedback control was suggested, where the output field of the cavity is directed to another quantum system (the controller) and then is fed back to the cavity without measurement; it was shown theoretically that this scheme can outperform measurement-based feedback, especially in the quantum regime \cite{Hamerly12,Jacobs14,Jacobs15}. In parallel with the progress in feedback cooling, open-loop control, where the interaction between the cavity field and the mirror is modulated by a predetermined function of time, was proposed as an alternative. In this context, numerical optimization methods \cite{Jacobs11,Machnes12} and optimized Lyapunov control \cite{Hou12} were employed to obtain control pulses which can in theory provide efficient cooling in shorter times than feedback (where cooling is achieved in steady state). The short cooling times are highly desirable to minimize the effect of the random interactions with the environment which ``heat" the mirror away from its ground state and towards thermal equilibrium.

Following the framework of open-loop control and after a brief presentation of the underlying physics in the next section, we formulate in section \ref{sec:problem} cavity optomechanical cooling as a minimum-time optimal control problem on anti-de Sitter space of appropriate dimension. The bounded control used in this approach is the coupling rate which modulates the interaction between the cavity field and the mechanical resonator. We also find the target point for this control problem using an approximation of the original system. In section \ref{sec:solution} we first show that the target point belongs indeed to the reachable set of the original system, and then use the Legendre pseudospectral optimization method \cite{Ross03} to find the minimum time and the corresponding optimal control to reach this point, for various values of the control bound. Finally, in the concluding section \ref{sec:conclusion}, we briefly discuss some further applications of the present work in quantum information processing and in the optimization of an optomechanical quantum heat engine.

\section{Cavity Optomechanical System}
\label{sec:system}

The system that we study in this paper consists of a mechanical resonator coupled to an optical resonator and is implemented experimentally with a laser driven optical cavity formed by two mirrors, one fixed and one movable which is attached to a spring \cite{Meystre13}. The light from the laser entering the cavity is reflected between the mirrors and the enhanced electromagnetic field built inside the cavity exerts a force to the movable mirror due to radiation pressure. The interaction between the cavity (photon) field and the mechanical vibration of the mirror (phonon field) can be described by the following Hamiltonian \cite{Aspelmeyer14,Jacobs11}
\begin{equation}\label{hamiltonianfull}
  \mathcal{H}=-\hbar\Delta\hat{a}^+\hat{a}+\hbar\omega_m\hat{b}^+\hat{b}+\hbar g(t)(\hat{a}^++\hat{a})(\hat{b}^++\hat{b}),
\end{equation}
where $\hbar$ is Planck's constant, $\Delta$ is roughly the laser detuning from the cavity frequency and $\omega_m$ is the frequency of the mechanical resonator. The operators $\hat{a},\hat{b}$ are the annihilation operators for the photon and the phonon fields, respectively, while $\hat{a}^+,\hat{b}^+$ are their hermitian adjoint, called creation operators \cite{Merzbacher97}. They satisfy the commutation relations
\begin{align}
[\hat{a},\hat{a}^+]&=[\hat{b},\hat{b}^+]=1,\nonumber\\
[\hat{a},\hat{b}^+]&=[\hat{a},\hat{b}]=0,
\end{align}
where the commutator of two operators $\hat{A}, \hat{B}$ is $[\hat{A},\hat{B}]=\hat{A}\hat{B}-\hat{B}\hat{A}$ \cite{Merzbacher97}.
The first two terms in (\ref{hamiltonianfull}) express the energy of each individual oscillator (optical and mechanical), while the last term the interaction between them. The coupling rate $g(t)$ has dimensions of frequency and is the available control which can be altered with time, with the aim to cool the mechanical resonator. To obtain a better understanding of the interaction term, we recall that the creation and annihilation operators are related to the position operators $\hat{x}_o,\hat{x}_m$ of the two oscillators through the relations $\hat{a}+\hat{a}^+\sim\hat{x}_o, \hat{b}+\hat{b}^+\sim\hat{x}_m$ \cite{Merzbacher97}, thus the interaction is proportional to the product $\hat{x}_o\hat{x}_m$. This is reminiscent of two classical oscillators with displacements $x_o, x_m$ which interact through the usual quadratic term $(x_o-x_m)^2$, from where the product $x_ox_m$ arises.

The state of the system can be described by the density matrix $\rho(t)$, which satisfies the Liouville-von Neumann equation \cite{Merzbacher97}
\begin{equation}\label{Liouvillefull}
  \dot{\rho}=-i[\mathcal{H}/\hbar,\rho].
\end{equation}
Note that here we consider only the coherent evolution and ignore relaxation, which describes the undesirable interaction of the system with its environment, since we are interested in the fast pulsed cooling and not the steady state feedback cooling. This is a legitimate practice in minimum-time quantum control problems \cite{Stefanatos11,Stefanatos12,Stefanatos13,Stefanatos14b}, the alternative being to include relaxation and maximize the fidelity of the final point \cite{Jacobs11,Stefanatos14a}. We will concentrate on the so-called ``red-detuned regime"
\begin{equation}\label{resonance}
  \Delta=-\omega_m
\end{equation}
where the two oscillators are on resonance. If we normalize time using the frequency $\omega_m$, i.e. set $t_{\mbox{new}}=\omega_mt_{\mbox{old}}$, then the Liouville-von Neumann equation becomes
\begin{equation}\label{Liouville}
  \dot{\rho}=-i[H,\rho]
\end{equation}
with the Hamiltonian
\begin{equation}\label{hamiltonian}
  H=\hat{a}^+\hat{a}+\hat{b}^+\hat{b}+g(t)(\hat{a}^++\hat{a})(\hat{b}^++\hat{b}),
\end{equation}
where the coupling rate is given in units of $\omega_m$. We can use (\ref{Liouville}) to derive ordinary differential equations for the expectation values of operators involved in the cooling of the mechanical resonator. For an operator $\hat{O}$ without explicit time-dependence, we have for the average value $O=\langle\hat{O}\rangle=\mbox{Tr}(\rho\hat{O})$ that \cite{Merzbacher97}
\begin{equation}\label{expectation}
  \dot{O}=d\langle\hat{O}\rangle/dt=i\langle[H,\hat{O}]\rangle.
\end{equation}
For the evolution described by (\ref{Liouville}) with Hamiltonian (\ref{hamiltonian}), a closed set of equations can be obtained for the operators formed by the second moments of the creation and annihilation operators of the two resonators, for example $\hat{a}^+\hat{a}, \hat{b}^+\hat{b}, \hat{a}^+\hat{b}, \hat{a}^2$ etc., see \cite{Jacobs11,Rae08}. In the next section we use specific linear combinations of these operators to formulate the cooling of the mechanical oscillator as a control problem. The benefits from this choice will immediately become evident there.

\section{Formulation of the Control Problem Using the Generators of the Symplectic Group $Sp(4)$}
\label{sec:problem}

In order to formulate the cooling of the mechanical resonator as a control problem, we will employ a set of ten operators which are the generators of the symplectic group $Sp(4)$ \cite{Han98}
\begin{align}\label{operators}
\hat{J}_0&=\frac{1}{2}(\hat{a}^+\hat{a}+\hat{b}\hat{b}^+),\quad\hat{J}_1=\frac{1}{2}(\hat{a}^+\hat{a}-\hat{b}^+\hat{b})\nonumber\\
\hat{J}_2&=\frac{1}{2}(\hat{a}^+\hat{b}+\hat{a}\hat{b}^+),\quad\hat{J}_3=\frac{1}{2i}(\hat{a}^+\hat{b}-\hat{a}\hat{b}^+)\nonumber\\
\hat{K}_1&=\frac{1}{2}(\hat{a}^+\hat{b}^++\hat{a}\hat{b}),\quad\hat{Q}_1=\frac{i}{2}(\hat{a}^+\hat{b}^+-\hat{a}\hat{b})\nonumber\\
\hat{K}_2&=-\frac{1}{4}[(\hat{a}^+)^2+\hat{a}^2-(\hat{b}^+)^2-\hat b^2],\nonumber\\
\hat{K}_3&=\frac{i}{4}[(\hat{a}^+)^2-\hat{a}^2+(\hat{b}^+)^2-\hat b^2],\nonumber\\
\hat{Q}_2&=-\frac{i}{4}[(\hat{a}^+)^2-\hat{a}^2-(\hat{b}^+)^2+\hat b^2],\nonumber\\
\hat{Q}_3&=-\frac{1}{4}[(\hat{a}^+)^2+\hat{a}^2+(\hat{b}^+)^2+\hat b^2].
\end{align}
Obviously, these operators are linear combinations of the second moments of the creation and annihilation operators of the two resonators and their matrix representation in terms of Pauli matrices can be found in \cite{Han98}. They satisfy the following commutation relations
\begin{align}\label{commutators}
  [\hat{J}_i,\hat{J}_j]&=i\epsilon_{ijk}\hat{J}_k,\quad [\hat{J}_i,\hat{K}_j]=i\epsilon_{ijk}\hat{K}_k,\nonumber\\
  [\hat{J}_i,\hat{Q}_j]&=i\epsilon_{ijk}\hat{Q}_k,\quad [\hat{K}_i,\hat{Q}_j]=i\delta_{ij}\hat{J}_0,\nonumber\\
  [\hat{K}_i,\hat{K}_j]&=[\hat{Q}_i,\hat{Q}_j]=-i\epsilon_{ijk}\hat{J}_k,\quad[\hat{J}_i,\hat{J}_0]=0,\nonumber\\
  [\hat{K}_i,\hat{J}_0]&=i\hat{Q}_i,\quad [\hat{Q}_i,\hat{J}_0]=-i\hat{K}_i,
\end{align}
where $\epsilon_{ijk}$ is the Levi-Civita symbol, which is $1$ if $(i,j,k)$ is an even permutation of $(1,2,3)$, $-1$ if it is an odd permutation, and $0$ if any index is repeated, while $\delta_{ij}$ is Kronecker's delta. In terms of operators (\ref{operators}), the optomechanical Hamiltonian (\ref{hamiltonian}) can be written as
\begin{equation}\label{hamiltonianfinal}
  H=2[\hat{J}_0-1/2+g(t)(\hat{K}_1+\hat{J}_2)].
\end{equation}
Using expression (\ref{hamiltonianfinal}) and the commutation relations (\ref{commutators}) in (\ref{expectation}), we obtain the following systems for the expectation values of the operators
\begin{equation}\label{4d}
  \left[\begin{array}{c}
    \dot{J}_1\\
    \dot{J}_3\\
    \dot{K}_2\\
    \dot{Q}_2
  \end{array}\right]
  =
  \left[\begin{array}{cccc}
    0 & 2g & 0 & 0 \\
    -2g & 0 & 2g & 0 \\
    0 & 2g & 0 & 2 \\
    0 & 0 & -2 & 0
  \end{array}\right]
  \left[\begin{array}{c}
    J_1\\
    J_3\\
    K_2\\
    Q_2
  \end{array}\right]
\end{equation}

\begin{equation}\label{6d}
  \left[\begin{array}{c}
    \dot{J}_0 \\
    \dot{Q}_1 \\
    \dot{K}_1 \\
    \dot{Q}_3 \\
    \dot{K}_3 \\
    \dot{J}_2
  \end{array}\right]
  =
  \left[\begin{array}{cccccc}
    0 & -2g & 0 & 0 & 0 & 0 \\
    -2g & 0 & -2 & 2g & 0 & 0 \\
    0 & 2 & 0 & 0 & 2g & 0 \\
    0 & -2g & 0 & 0 & -2 & 0 \\
    0 & 0 & -2g & 2 & 0 & -2g \\
    0 & 0 & 0 & 0 & -2g & 0
  \end{array}\right]
  \left[\begin{array}{c}
    J_0 \\
    Q_1 \\
    K_1 \\
    Q_3 \\
    K_3 \\
    J_2
  \end{array}\right]
\end{equation}

We next move to express the initial and target states of the above systems. At $t=0$ we consider the situation where there is a nonzero average phonon number $\langle\hat{b}^+\hat{b}\rangle=n_b>0$ in the mechanical vibrator, while the number of photons is zero $\langle\hat{a}^+\hat{a}\rangle=0$, an approximation which is valid for optical fields at room temperature \cite{Aspelmeyer14}. The initial values of the rest of the second moments are also taken to be zero. We obtain the following initial conditions for the expectation values of the operators defined in (\ref{operators})
\begin{align}
\left[\begin{array}{cccc}
    J_1 & J_3 & K_2 & Q_2
\end{array}\right]
&=
\left[\begin{array}{cccc}
    -\frac{n_b}{2} & 0 & 0 & 0
\end{array}\right]\nonumber\\
\left[\begin{array}{cccccc}
    J_0 & Q_1 & K_1 & Q_3 & K_3 & J_2
\end{array}\right]
&=
\left[\begin{array}{cccccc}
    \frac{n_b+1}{2} & 0 & 0 & 0 & 0 & 0
\end{array}\right].\nonumber
\end{align}

The cooling of the mechanical resonator corresponds to the application of the appropriate control $g(t)$ which minimizes the average number of phonons $\langle\hat{b}^+\hat{b}\rangle$ at the final state. In order to obtain a hint about this target state, we examine the limiting case of constant $g\ll 1$, where the so called rotating wave approximation (RWA) is valid \cite{Liu14}. Under this approximation, the part $2g\hat{K}_1$ of the interaction term in (\ref{hamiltonianfinal}) can be neglected and the Hamiltonian becomes $H_{\mbox{RWA}}=2(\hat{J}_0-1/2+g\hat{J}_2)$. Observe from (\ref{commutators}) that $\hat{J}_0$ commutes with $\hat{J}_i, i=1,2,3$ and thus with $H_{\mbox{RWA}}$, so the expectation value $J_0$ is constant, while $\hat{J}_i, i=1,2,3$ satisfy the spin commutation relations. From this analogy we infer that the evolution under $H_{\mbox{RWA}}$ is actually a rotation of the vector $\mathbf{J}=[J_1\,J_2\,J_3]^T$ around the 2-axis. In order to see this clearly we use $H_{\mbox{RWA}}$ in (\ref{expectation}) and find $\dot{J}_1=2gJ_3$, $\dot{J}_2=0$ and $\dot{J}_3=-2gJ_1$. Using these equations and the above initial conditions we obtain the constants of the motion $J_1^2+J_2^2+J_3^2=(n_b/2)^2$ and $J_2=0$. It is now clear that the vector $\mathbf{J}$ is restricted on a sphere and is rotated around the 2-axis with angular velocity $2g$, starting from the point $\mathbf{J}(0)=[-\frac{n_b}{2}\,0\,0]^T$. From (\ref{operators}) it is obvious that the difference $\langle\hat{a}^+\hat{a}\rangle-\langle\hat{b}^+\hat{b}\rangle$ between the average photon and phonon numbers is maximized when $J_1$ is maximum, which occurs when the antipodal point $\mathbf{J}(\frac{\pi}{2g})=[\frac{n_b}{2}\,0\,0]^T$ is reached. But the sum $\langle\hat{a}^+\hat{a}\rangle+\langle\hat{b}^+\hat{b}\rangle$ is constant since $J_0$ is constant. Thus, when the antipodal point is reached, the number of phonons is actually minimized while the number of photons is maximized, with corresponding values $\langle\hat{b}^+\hat{b}\rangle=0$ and $\langle\hat{a}^+\hat{a}\rangle=n_b$. Obviously, the photon-phonon populations have been swapped. It is not hard to see that the values of the rest of the second moments are zero at the antipodal point.

Based on the above observations for the case where the rotating wave approximation holds, we move to the original system described by the full Hamiltonian (\ref{hamiltonianfinal}) and require the same target point at the final time $t=T$. In terms of the operators (\ref{operators}), the desired final point can be expressed as
\begin{align}
\left[\begin{array}{cccc}
    J_1 & J_3 & K_2 & Q_2
\end{array}\right]
&=
\left[\begin{array}{cccc}
    \frac{n_b}{2} & 0 & 0 & 0
\end{array}\right]\nonumber\\
\left[\begin{array}{cccccc}
    J_0 & Q_1 & K_1 & Q_3 & K_3 & J_2
\end{array}\right]
&=
\left[\begin{array}{cccccc}
    \frac{n_b+1}{2} & 0 & 0 & 0 & 0 & 0
\end{array}\right].\nonumber
\end{align}
Observe that for subsystem (\ref{4d}) this target point is the antipodal of the initial point, while for subsystem (\ref{6d}) the two points coincide.

If we normalize the expectation values $J_1, J_3, K_2, Q_2$ with $n_b/2$ and the expectation values $J_0, Q_1, K_1, Q_3, K_3, J_2$ with $(n_b+1)/2$, but keep the same notation for these variables, then systems (\ref{4d}) and (\ref{6d}) remain unchanged, while the starting and target points become
\begin{align}\label{start}
\left[\begin{array}{cccc}
    J_1 & J_3 & K_2 & Q_2
\end{array}\right]
&=
\left[\begin{array}{cccc}
    -1 & 0 & 0 & 0
\end{array}\right]\nonumber\\
\left[\begin{array}{cccccc}
    J_0 & Q_1 & K_1 & Q_3 & K_3 & J_2
\end{array}\right]
&=
\left[\begin{array}{cccccc}
    1 & 0 & 0 & 0 & 0 & 0
\end{array}\right]
\end{align}
and
\begin{align}\label{target}
\left[\begin{array}{cccc}
    J_1 & J_3 & K_2 & Q_2
\end{array}\right]
&=
\left[\begin{array}{cccc}
    1 & 0 & 0 & 0
\end{array}\right]\nonumber\\
\left[\begin{array}{cccccc}
    J_0 & Q_1 & K_1 & Q_3 & K_3 & J_2
\end{array}\right]
&=
\left[\begin{array}{cccccc}
    1 & 0 & 0 & 0 & 0 & 0
\end{array}\right],
\end{align}
independent of the initial phonon population $n_b$. Using the initial conditions (\ref{start}) and Eqs. (\ref{4d}) and (\ref{6d}), the following constants of the motion can be easily verified
\begin{align}\label{antidesitter}
K_2^2+Q_2^2-J_1^2-J_3^2 &= -1 \\
K_1^2+Q_1^2+K_3^2+Q_3^2-J_0^2-J_2^2 &= -1
\end{align}
Thus, systems (\ref{4d}) and (\ref{6d}) evolve on the anti-de Sitter spaces $AdS_3$ and $AdS_5$, respectively \cite{Bengsson98}.

Now we can formulate the cavity optomechanical cooling as a time-optimal control problem. We would like to find the bounded control $-G_0\leq g(t)\leq G_0$, $G_0>0$ , which drives systems (\ref{4d}) and (\ref{6d}) from the starting point (\ref{start}) to the target point (\ref{target}), in minimum time. Before doing so, we show first that the target point (\ref{target}), obtained using the rotating wave approximation, belongs indeed to the reachable set of the full system, for every control upper bound $G_0>0$.

\section{Reachability of the Target Point and Numerical Solution of the Minimum-Time Problem}
\label{sec:solution}

\begin{thm}\label{main}
The target point (\ref{target}) belongs to the reachable set of the full system (\ref{4d}), (\ref{6d}) with $-G_0\leq g(t)\leq G_0$, for every control upper bound $G_0>0$.
\end{thm}
\begin{pf*}{Proof.}
The symplectic group $Sp(4)$ is not compact, thus we cannot apply the Lie algebraic tools, described for example in \cite{Brockett14}, to prove the stronger result of controllability. Instead, we will show the reachability of the target point following a straightforward approach. For a specified upper bound $G_0>0$ consider a constant control $g(t)=G$ with $0<G\leq\mbox{min}\{G_0, 1/2\}$. In this case, linear system (\ref{4d}) has imaginary eigenvalues $\lambda=\pm i(\omega_+\pm\omega_-)$, with
\begin{equation}\label{omega}
  \omega_\pm=\sqrt{1\pm 2G},
\end{equation}
while linear system (\ref{6d}) has the eigenvalue $\lambda=0$ with double multiplicity and the imaginary eigenvalues $\lambda=\pm i2\omega_\pm$. For a time evolution of duration $t=T$ the state of (\ref{4d}) can be expressed as a linear combination of $\sin[(\omega_+\pm\omega_-)T]$ and $\cos[(\omega_+\pm\omega_-)T]$. The target point for this system is the antipodal of the starting point, thus it can be reached when
\begin{equation}\label{condition}
  (\omega_++\omega_-)T=m\pi,\quad (\omega_+-\omega_-)T=n\pi,
\end{equation}
where $m, n$ are positive \textbf{odd integers} with $m>n$. For system (\ref{6d}), the zero double eigenvalue has two independent eigenvectors, thus the state of this system can be expressed as a linear combination of constant terms and $\sin(2\omega_\pm T), \cos(2\omega_\pm T)$. The target point for this system is the same as the starting point and this requirement can be fulfilled if $2\omega_\pm T$ are even multiples of $\pi$, which is automatically satisfied when (\ref{condition}) holds with $m, n$ odd. From (\ref{omega}), (\ref{condition}) we find
\begin{equation}\label{T}
  T=\frac{\pi}{2}\sqrt{m^2+n^2}
\end{equation}
and
\begin{equation}\label{G}
  G=\frac{mn}{m^2+n^2},
\end{equation}
similar to the relations obtained in \cite{Liu14}. To complete the proof we show that a constant control $G$ of the form (\ref{G}), satisfying the constraint $G\leq G_0$, can be found for every bound $G_0>0$. If $G_0\geq 1/2$ then every pair $(m,n)$ with $m>n$ satisfies $G<G_0$. If $G_0<1/2$, the requirement $G\leq G_0$ leads to
\begin{equation}\label{fraction_bound}
  \frac{n}{m}\leq\frac{1-\sqrt{1-4G_0^2}}{2G_0}< 1,
\end{equation}
which can be satisfied with the appropriate choice of $(m,n)$ with $m>n$.\qed
\end{pf*}
\begin{cor}
The minimum necessary time $T$ to reach the target point with constant control, when the control bound is restricted as
\begin{equation}\label{pbound}
  \frac{m}{m^2+1}\leq G_0< \frac{m-2}{(m-2)^2+1},\quad m=3, 5, 7,\ldots
\end{equation}
is
\begin{equation}\label{Tp}
  T=\frac{\pi}{2}\sqrt{m^2+1}.
\end{equation}
\end{cor}
\begin{pf*}{Proof.}
Suppose that the minimum time is achieved for a pair $(m,n)$ with $m,n$ positive odd integers and $m>n>1$. The fraction $n/m$ satisfies condition (\ref{fraction_bound}) and so does the pair $(m,1)$ since $1/m<n/m$. But the time (\ref{Tp}) corresponding to $(m,1)$ is less than the time (\ref{T}) corresponding to $(m,n)$, thus the assumption that the minimum time is obtained for $n>1$ is wrong. The minimum time is achieved for pairs of the form $(m,1)$ and the corresponding constant control is $G=m/(m^2+1)$, when the control bound $G_0$ is restricted as in (\ref{pbound}).\qed
\end{pf*}

Having established the reachability of the target point, we next move to find the minimum necessary time to reach it for various values of the control bound $G_0$. Although analytical solutions have been obtained for a minimum-energy problem on the non-compact group $SU(1,1)$ (generated by $\hat{J}_0, \hat{K}_1, \hat{Q}_1$) \cite{Dong15}, solving analytically the minimum-time control problem at hand seems to be a formidable task due to its large dimension, thus here we recourse to numerical optimization and use the Legendre pseudospectral method \cite{Ross03}. This method has been extensively used in aerospace applications for trajectory optimization \cite{Bedrossian12}, but we have also used it in the control of quantum systems. We mention for example the pulse design in nuclear magnetic resonance (NMR) spectroscopy \cite{Li09,Li11}, while recently we used it to maximize the performance of a quantum heat engine in the presence of external noise \cite{Stefanatos14a}. The idea behind the method is to discretize the time interval using a fixed set of not equally spaced nodes. The states and controls are approximated by polynomials with exact values on these interpolation nodes, while the spacing between the nodes is such that the approximation error for other points is close to minimum. Using this approximation, the continuous-time dynamics described by ordinary differential equations is converted to a set of algebraic equations for the values of the states and controls on the nodes. The optimal control problem is thus transformed to a discrete nonlinear programming problem, which can be solved by many well-developed computational algorithms. More details of the method can be found in \cite{Li09,Stefanatos14a} and will not be repeated here.

\begin{figure}[t]
\begin{center}
\includegraphics[width=0.9\linewidth]{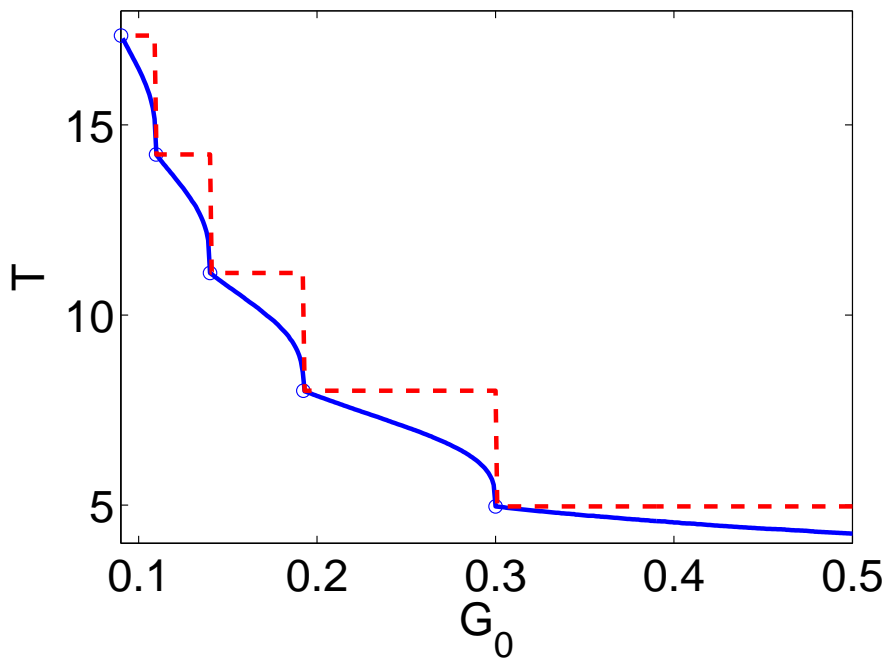}    
\caption{Numerically obtained minimum time values $T$ (blue solid line, units $\omega^{-1}_m$) for control bound in the interval $11/122\leq G_0\leq 1/2$ (units $\omega_m$). The circles indicate the points where the minimum-time control is a bang pulse.}                         
\label{fig:time}                           
\end{center}                               
\end{figure}

\begin{figure*}[t]
 \centering
		\begin{tabular}{cc}
     	\subfigure[$\ $$G_0=0.22$ (units $\omega_m$)]{
	            \label{fig:control1}
	            \includegraphics[width=.4\linewidth]{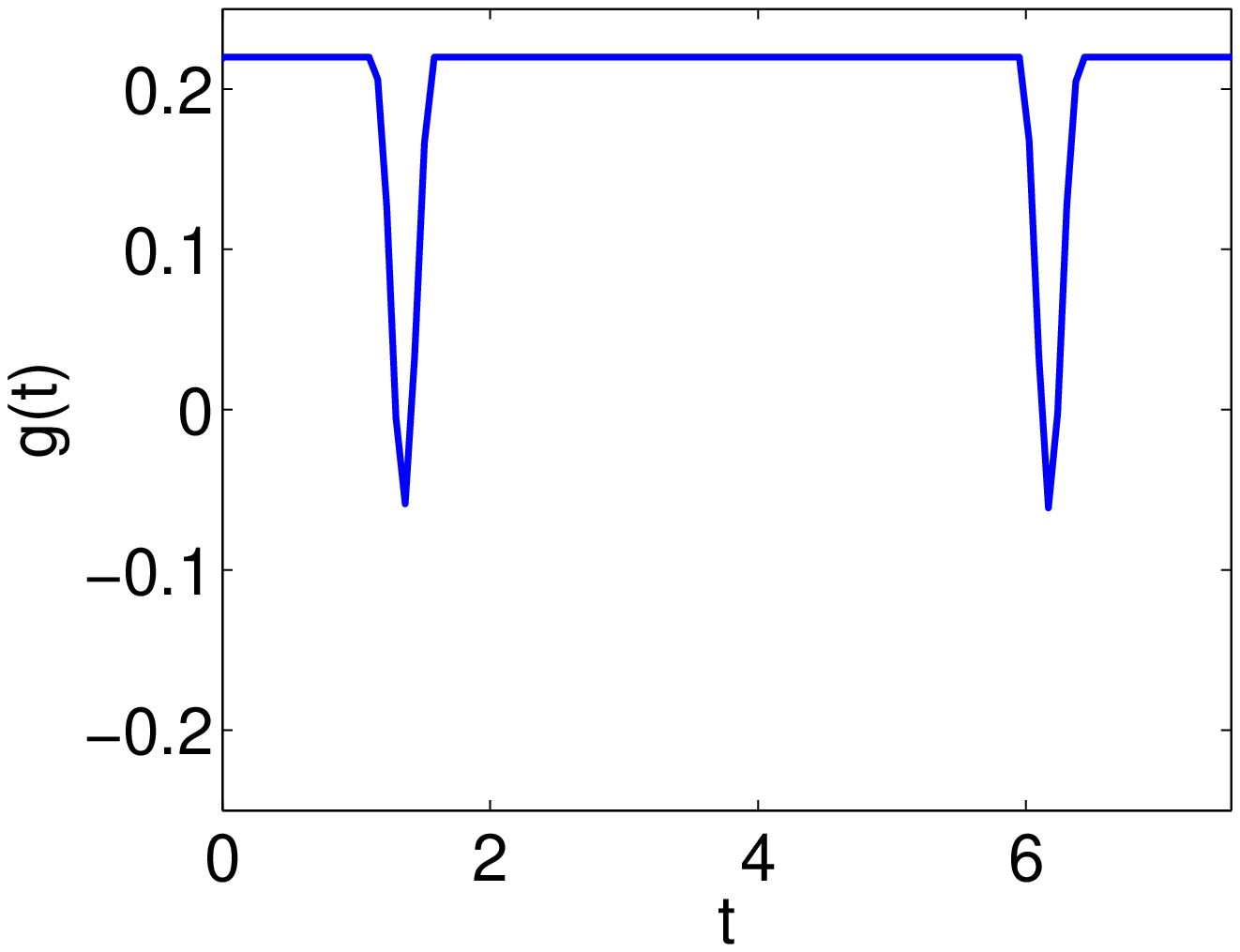}} &
	        \subfigure[$\ $ $G_0=0.22$ (units $\omega_m$)]{
	            \label{fig:evolution1}
	            \includegraphics[width=.4\linewidth]{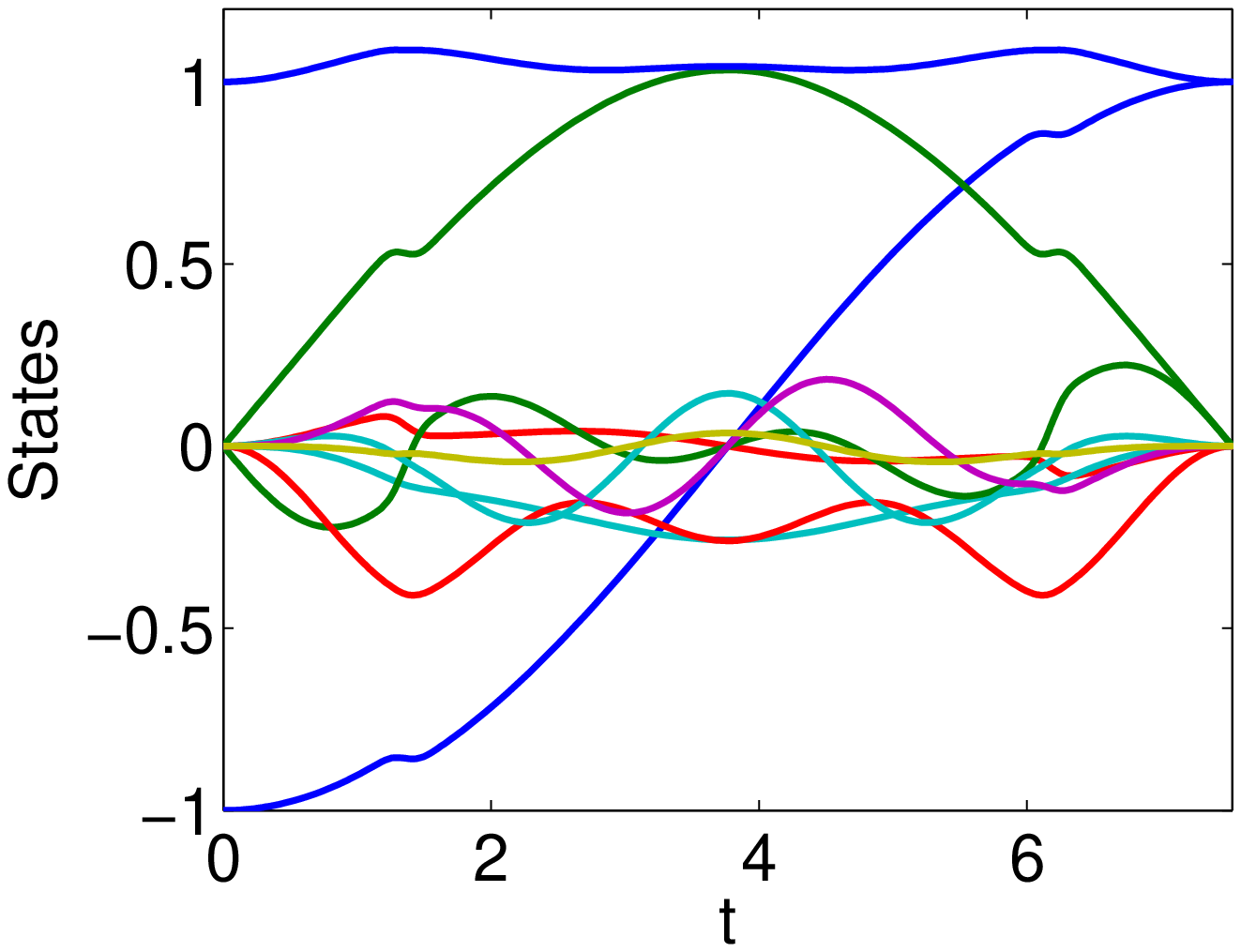}} \\
	        \subfigure[$\ $$G_0=0.50$ (units $\omega_m$)]{
	            \label{fig:control2}
	            \includegraphics[width=.4\linewidth]{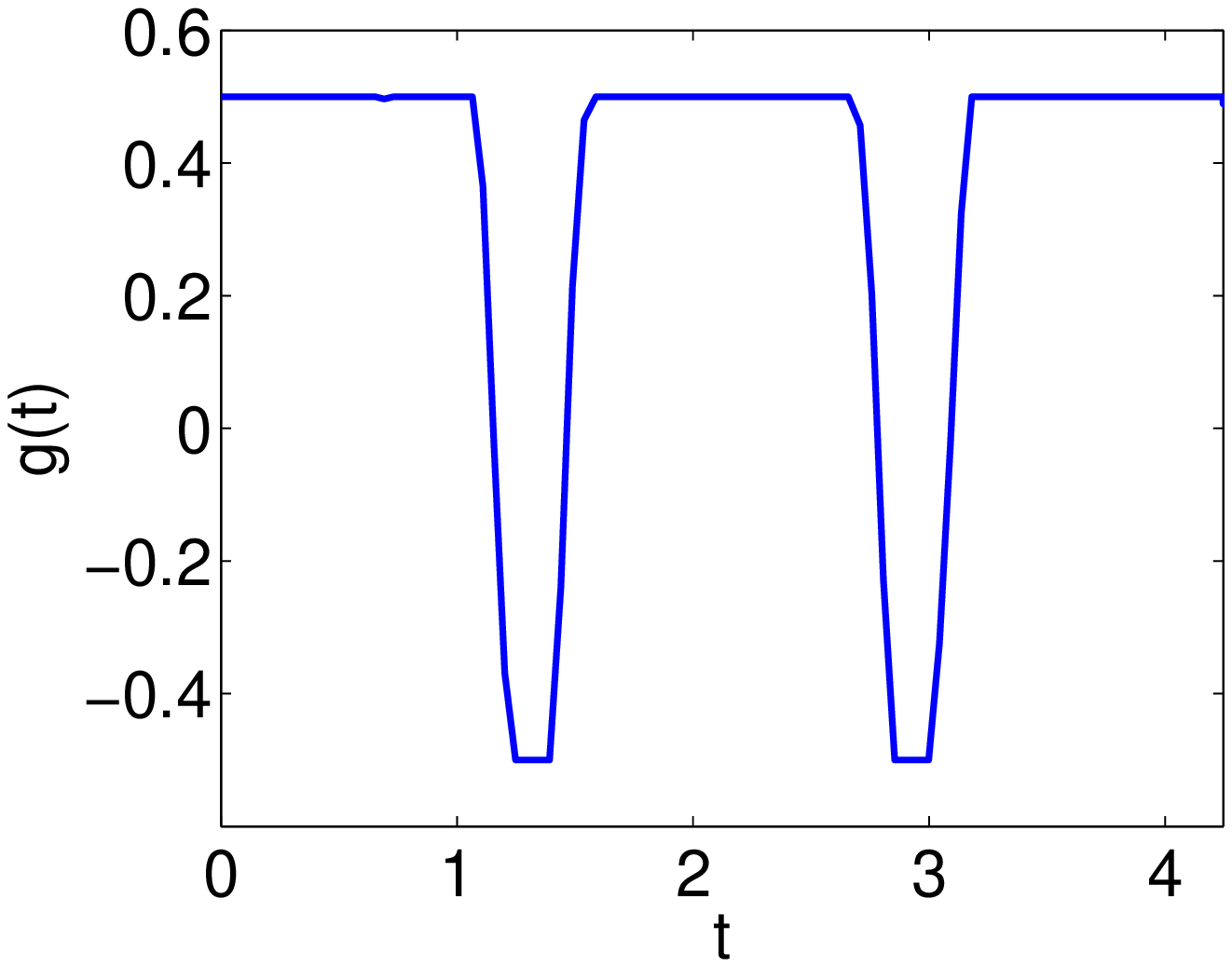}} &
			\subfigure[$\ $$G_0=0.50$ (units $\omega_m$)]{
	            \label{fig:evolution2}
	            \includegraphics[width=.4\linewidth]{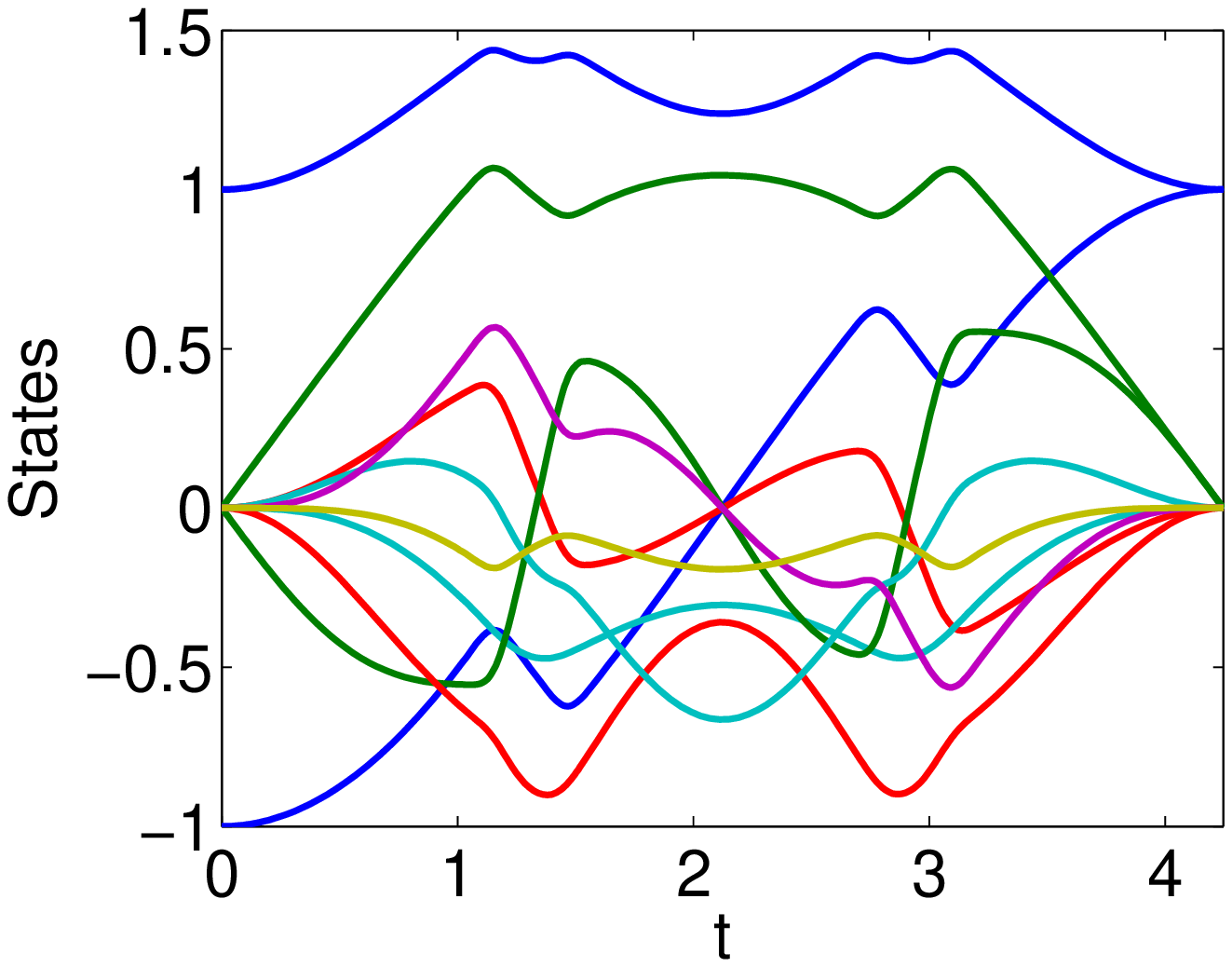}}
		\end{tabular}
\caption{Optimal control $g(t)$ (units $\omega_m$) and corresponding state evolution for two values of $G_0$. Time is given in units of $\omega^{-1}_m$.}
 \label{fig:sim}
\end{figure*}

In Fig. \ref{fig:time} we plot the minimum time $T$ (blue solid line) for various values of the control bound in the interval $11/122\leq G_0\leq 1/2$. In order to obtain this plot we solved numerically a series of optimization problems using $N=69+1=70$ interpolation nodes. Starting from the larger value of the bound $G_0=1/2$ and with time initially restricted in the interval $4.00\leq T\leq 4.10$, we run the Legendre pseudospectral method searching for the minimum time $T$ which satisfies the discretized system dynamics and the boundary conditions. We implement the method using AMPL (A Mathematical Programming Language) with MINOS $5.51$ solver \cite{AMPL}. If the problem is infeasible, we increase the upper bound for time by $0.001$ and repeat the optimization. If a minimum time $T$ is found, we set the lower bound for time to this optimal value, we increase the upper bound for time by $0.001$, we reduce $G_0$ by $\Delta G_0=-0.001$, and repeat the same procedure all the way down to $G_0=11/122$. These steps are actually codified in an AMPL script. Note that the above described procedure relies on the fact that the minimum time $T$ is a monotonically decreasing function of the upper bound $G_0$. In the same figure we also plot the times from (\ref{Tp}) for $m=3,5,7,9,11$ (blue circles), where the target point is reached with constant (bang) control $g(t)=G_0$. The numerical solution indicates that these times are optimal for the corresponding values of $G_0=m/(m^2+1)$. Observe that these times form a staircase with steps of increasing height (tending to the limiting value $\pi$) and decreasing width (tending to zero); the first five steps are depicted in Fig. \ref{fig:time} (red dashed line). Finally observe that at these points, where the optimal control becomes a single bang pulse, the slope of the minimum time curve is discontinuous, a similar behavior to the one we have encountered in another minimum-time problem, see Fig. 4 in \cite{Stefanatos14b}.

In Fig. \ref{fig:sim} we plot the optimal control $g(t)$ and the corresponding evolution of systems (\ref{4d}) and (\ref{6d}) for two values of the control bound, $G_0=0.22$ (Figs. \ref{fig:control1}, \ref{fig:evolution1}) and $G_0=0.50$ (Figs. \ref{fig:control2}, \ref{fig:evolution2}), using $129+1=130$ interpolation nodes in the corresponding simulations for better resolution. Observe that for the larger control bound, which corresponds to a shorter transfer time, the optimal control approaches a bang-bang form, taking most of its values on the boundaries. Also note that in both cases the optimal control is symmetric. It is not hard to see that this symmetry is actually a general characteristic of the optimal control. Let $g(t)$ be the optimal control for the minimum-time problem stated in section \ref{sec:problem}. If we consider the backward evolution with $\sigma=T-t$ and also make the change in the state variables $x\rightarrow -x$, we recover equations (\ref{4d}) and (\ref{6d}). The initial and final points for (\ref{4d}) are the same as in the forward evolution, while the initial and final point for (\ref{6d}) is now $\left[\begin{array}{cccccc}-1 & 0 & 0 & 0 & 0 & 0\end{array}\right]$. The optimal control which drives (\ref{4d}) to the target point is obviously $g(\sigma)=g(T-t)$, and we know from the forward evolution that this control, when applied to (\ref{6d}), returns the initial point $\left[\begin{array}{cccccc} 1 & 0 & 0 & 0 & 0 & 0\end{array}\right]$ to itself. Because of the linearity of system (\ref{6d}) we conclude that the same control returns also the initial point $\left[\begin{array}{cccccc}-1 & 0 & 0 & 0 & 0 & 0\end{array}\right]$ of the backward evolution to itself. Thus $g(T-t)$ is the optimal control for the backward evolution. But the minimum-time control for the forward and the backward evolution is the same, so $g(t)=g(T-t)$, and this is the symmetry depicted in Figs. \ref{fig:control1} and \ref{fig:control2}.

We close the analysis by making some observations about the state evolution shown in Figs. \ref{fig:evolution1} and \ref{fig:evolution2}. In both pictures, we can easily identify $J_1$ (the blue solid line connecting $-1$ to $1$), $J_0$ (the blue solid line starting and ending to $1$), and $J_3$ (the green solid line which is the largest among the rest of the states). For the case of the smaller control bound, depicted in Fig. \ref{fig:evolution1}, the evolution of $J_1$ and $J_3$ is close to sinusoidal, $J_0\approx 1$ throughout, while the rest of the states are substantially smaller. For the case of the larger control bound, shown in Fig. \ref{fig:evolution2}, $J_1, J_3$ deviate from the sinusoidal behavior, $J_0$ is not approximately constant anymore, while the rest of the states have larger values than before and actively participate in the transfer. This difference can be explained from the fact that in the former case the transfer time is larger (the situation is more ``adiabatic") and thus RWA provides a fair description, while in the latter case the shorter transfer time ``excites" the path neglected by the RWA.

\section{Conclusion and Future Work}
\label{sec:conclusion}

In this paper we formulated cavity optomechanical cooling as a minimum-time control problem and we used the Legendre pseudospectral method to obtain numerically  the optimal control, for various values of the coupling strength between the cavity field and the movable mirror. The advantages of this numerical optimization method are its simplicity and the ability to easily incorporate constraints \cite{Li09,Stefanatos14a}. Thus, by simply changing the target point, we can immediately use this method for minimum-time entanglement creation between the movable mirror and the cavity field \cite{Vitali07}. Additionally, if we use as a control the detuning $\Delta$ of the driving laser, instead of the coupling rate, then similar methodology can be applied for speeding up the adiabatic-like stroke of a recently suggested optomechanical quantum heat engine, which is a good candidate for practical implementation \cite{Zhang14}. All these applications are examples where quantum control can provide the protocols for realizing the operation of quantum devices within their performance limits, as pointed out in the recently published ``roadmap to quantum control" \cite{Glaser15}.



\begin{ack}
The author would like to thank the anonymous referee of \cite{Stefanatos14a}, for bringing to his attention the problem of evaluating the limits of cavity optomechanical cooling, and AMPL Optimization Inc., for providing a free trial licence of AMPL and MINOS solver.
\end{ack}


\begin{thebibliography}{99}     

\balance

\bibitem{Aspelmeyer14}
Aspelmeyer, M., Kippenberg, T.J., \& Marquardt, F. (2014). Cavity optomechanics.
{\it Reviews of Modern Physics, 86}, 1391.

\bibitem{Bedrossian12}
Bedrossian, N., Karpenko, M., \& Bhatt, S. (2012). Overclock my satellite.
{\it IEEE Spectrum, 49}(11), 54-62.

\bibitem{Bengsson98}
Bengsson, I. (1998). {\it Anti-de Sitter Space}. Lecture notes, Stockholm University, Sweden. 

\bibitem{Brockett14}
Brockett, R. (2014). The early days of geometric nonlinear control. {\it Automatica, 50}(9), 2203–2224.

\bibitem{Cohadon99}
Cohadon, P.F., Heidmann, A., \& Pinard, M. (1999). Cooling a mirror by radiation pressure.
{\it Physical Review Letters, 83}, 3174.

\bibitem{Doherty99}
Doherty, A.C., \& Jacobs, K. (1999). Feedback control of quantum systems using continuous state estimation.
{\it Physical Review A, 60}, 2700.

\bibitem{Dong10}
Dong, D., \& Petersen, I.R. (2010). Quantum control theory and applications: A survey. {\it IET Control Theory \& Applications, 4}, 2651-2671.

\bibitem{Dong15}
Dong, W., Wu, R., Wu, J., Li, C., \& Tzyh-Jong Tarn, T.J (2015). Optimal control of quantum systems with $SU(1,1)$ dynamical symmetry. {\it Control Theory \& Technology, 13}(3), 211–220.

\bibitem{AMPL}
Fourer, R., Gay D.M, \& Kernighan, B.W. (2002). {\it AMPL: A Modeling Language for Mathematical Programming}. Duxbury Press.

\bibitem{Glaser15}
Glaser, S.J., et al. (2015). Training Schrödinger's cat: quantum optimal control. {\it arXiv}, quant-ph/1508.00442.

\bibitem{Gough09}
Gough, J., \& James, M.R. (2009). The series product and its application to quantum feedforward and feedback networks.
{\it IEEE Transactions on Automatic Control, 54}(11), 2530-2544.

\bibitem{Hamerly12}
Hamerly, R., \& Mabuchi, H. (2012). Advantages of coherent feedback for cooling quantum oscillators.
{\it Physical Review Letters, 109}, 173602.

\bibitem{Han98}
Han, D., Kim, Y.S. (1998). Squeezed states as representations of symplectic groups. {\it arXiv}, physics/9803017.

\bibitem{Hou12}
Hou, S.C., Khan, M.A., Yi, X.X., Dong, D., \& Petersen, I.R. (2012). Optimal Lyapunov-based quantum control for quantum systems. {\it Physical Review A, 86}, 022321.

\bibitem{Jacobs15}
Jacobs, K., Nurdin, H.I., Strauch, F.W., \& James, M. (2015). Comparing resolved-sideband cooling and measurement-based feedback cooling on an equal footing: Analytical results in the regime of ground-state cooling. {\it Physical Review A, 91}, 043812.

\bibitem{Jacobs14}
Jacobs, K., Wang, X., \& Wiseman, H.M. (2014). Coherent feedback that beats all measurement-based feedback protocols.
{\it New Journal of Physics, 16}, 073036.

\bibitem{James08}
James, M.R., Nurdin, H., \& Petersen, I.R. (2008). Control of linear quantum stochastic systems.
{\it IEEE Transactions on Automatic Control, 53}(8), 1787-1803.

\bibitem{Li09}
Li, J.-S., Ruths, J., \& Stefanatos, D. (2009). A pseudospectral method for optimal control of open quantum systems. {\it Journal of Chemical Physics, 131}, 164110.

\bibitem{Li11}
Li, J.-S., Ruths, J., Yu, T.Y., Arthanari, H., \& Wagner, G. (2011). Optimal pulse design in quantum control: A unified computational method. {\it Proceedings of National Academy of Sciences U.S.A., 108}(5), 1879-1884.

\bibitem{Liu13}
Liu, Y.C., Hu, Y.W., Wong, C.W., \& Xiao, Y.F. (2013). Review of cavity optomechanical cooling.
{\it Chinese Physics B, 22}(11), 114213.

\bibitem{Liu14}
Liu, Y.C., Shen, Y.F., Gong, Q., \& Xiao, Y.F. (2014). Optimal limits of cavity optomechanical cooling in the strong-coupling regime.
{\it Physical Review A, 89}, 053821.

\bibitem{Machnes12}
Machnes, S., Cerrillo, J., Aspelmeyer, M., Wieczorek, W., Plenio, M.B., \& Retzker, A. (2012). Pulsed laser cooling for cavity optomechanical resonators.
{\it Physical Review Letters, 108}, 153601.

\bibitem{Mancini98}
Mancini, S., Vitali, D., \& Tombesi, P. (1998). Optomechanical cooling of a macroscopic oscillator by homodyne feedback.
{\it Physical Review Letters, 80}, 688.

\bibitem{Merzbacher97}
Merzbacher, E. (1997). {\it Quantum Mechanics}. Wiley.

\bibitem{Meystre13}
Meystre, P. (2013). A short walk through quantum optomechanics.
{\it Annalen der Physik (Berlin), 525}, 215–233.

\bibitem{Mori09}
Mori, O., et al. (2009). Development of first solar power sail demonstrator - IKAROS. {\it Proceedings of the International Symposium on Space Flight Dynamics}, Toulouse, France.

\bibitem{Nurdin09}
Nurdin, H., James, M.R., \& Petersen, I.R. (2009). Coherent quantum LQG control.
{\it Automatica, 45}(8), 1837-1846.

\bibitem{Ross03}
Ross, I., \& Fahroo, F. (2003). {\it Legendre Pseudospectral Approximations of Optimal Control Problems}. Lecture Notes in Control and Information Sciences 295, Berlin: Springer.

\bibitem{Stefanatos14a}
Stefanatos, D. (2014). Optimal efficiency of a noisy quantum heat engine.
{\it Physical Review E, 90}, 012119.

\bibitem{Stefanatos13}
Stefanatos, D. (2013). Optimal shortcuts to adiabaticity for a quantum piston.
{\it Automatica, 49}(10), 3079-3083.

\bibitem{Stefanatos12}
Stefanatos, D., \& Li, J.-S. (2012). Frictionless decompression in minimum time of Bose-Einstein condensates in the Thomas-Fermi regime.
{\it Physical Review A, 86}, 063602.

\bibitem{Stefanatos14b}
Stefanatos, D., \& Li, J.-S. (2014). Minimum-time quantum transport with bounded trap velocity.
{\it IEEE Transactions on Automatic Control, 59}(3), 733-738.

\bibitem{Stefanatos11}
Stefanatos, D., Schaettler, H., \& Li, J.-S. (2011). Minimum-time frictionless atom cooling in harmonic traps.
{\it SIAM Journal on Control and Optimization, 49}(6), 2440-2462.

\bibitem{Vitali07}
Vitali, D., Gigan, S., Ferreira, A., B\"{o}hm, H.R., Tombesi, P., Guerreiro, A., Vedral, V., Zeilinger, A., \& Aspelmeyer, M. (2007). Optomechanical entanglement between a movable mirror and a cavity field.
{\it Physical Review Letters, 98}, 030405.

\bibitem{Jacobs11}
Wang, X., Vinjanampathy, S., Strauch, F.W., \& Jacobs, K. (2011). Ultraefficient cooling of resonators: beating sideband cooling with quantum control.
{\it Physical Review Letters, 107}, 177204.

\bibitem{Wieczorek15}
Wieczorek, W., Hofer, S.G., Hoelscher-Obermaier, J., Riedinger, R., Hammerer, K., \& Aspelmeyer, M. (2015). Optimal state estimation for cavity optomechanical systems.
{\it Physical Review Letters, 114}, 223601.

\bibitem{Rae08}
Wilson-Rae, I., Nooshi, N., Dobrindt, J., Kippenberg, T.J., \& Zwerger, W. (2008). Cavity-assisted backaction cooling of mechanical resonators.
{\it New Journal of Physics, 10}, 095007.

\bibitem{Zhang14}
Zhang, K., Bariani, F., \& Meystre, P. (2014). Quantum optomechanical heat engine.
{\it Physical Review Letters, 112}, 150602.

















\end{thebibliography}


\end{document}